# Fast Charging of Lithium-Ion Batteries Using Deep Bayesian Optimization with Recurrent Neural Network

Benben Jiang, *Member, IEEE*, Yixing Wang, Zhenghua Ma, and Qiugang Lu

*Abstract*—Fast charging has attracted increasing attention from the battery community for electrical vehicles (EVs) to alleviate range anxiety and reduce charging time for EVs. However, inappropriate charging strategies would cause severe degradation of batteries or even hazardous accidents. To optimize fast-charging strategies under various constraints, particularly safety limits, we propose a novel deep Bayesian optimization (BO) approach that utilizes Bayesian recurrent neural network (BRNN) as the surrogate model, given its capability in handling sequential data. In addition, a combined acquisition function of expected improvement (EI) and upper confidence bound (UCB) is developed to better balance the exploitation and exploration. The effectiveness of the proposed approach is demonstrated on the PETLION, a porous electrode theory-based battery simulator. Our method is also compared with the state-of-the-art BO methods that use Gaussian process (GP) and non-recurrent network as surrogate models. The results verify the superior performance of the proposed fast charging approaches, which mainly results from that: (i) the BRNN-based surrogate model provides a more precise prediction of battery lifetime than that based on GP or non-recurrent network; and (ii) the combined acquisition function outperforms traditional EI or UCB criteria in exploring the optimal charging protocol that maintains the longest battery lifetime.

## I. INTRODUCTION

Fast charging is an essential technology for alleviating the issues of mileage anxiety and overly long charging time for electrical vehicles (EVs), and thus it has drawn increasing attention in recent years. However, fast charging with extremely high power may intensify side reactions such as lithium dendrite growth and electrolyte decomposition[1] that can considerably degrade batteries. Fast charging process could also produce excessive heat which may lead to safety risks such as fire and explosion[2][3]. Therefore, it is imperative to develop fast charging strategies that can maintain the safe and efficient operation of batteries while minimizing the battery degradation.

The optimization of charging strategy has received broad attention due to the widespread demand for fast charging in various daily-life and industrial applications. In general, the optimization mainly focuses on the selection of charging currents, aiming to reduce battery's degradation and maintain battery's safe operation to the maximum extent. Existing methods on this topic can be roughly classified into model-based and data-driven approaches. For model-based methods, commonly employed battery models include equivalent circuit model (ECM)[4], single particle model (SPM)[5], and pseudo-two-dimensional model (P2D)[6]. In particular, a single shooting method with a coupled electrochemical-thermal aging model has been proposed to design a fast-charging strategy[7], where the growth of the solid electrolyte interface (SEI) layer can be captured. Ashwin et al.[8] modeled battery aging mechanism using the P2D model, accounting for the thickness of SEI layer and the temperature. Parhizi et al.[9] studied the capacity degradation of lithium-ion batteries under different operating conditions by analyzing the growth of SEI layer based on the SPM. Xavier et al.[10] developed model-predictive control (MPC) based on reduced-order models to design fast charging strategies under the constraint of no Li plating. A quadratic dynamic matrix control method is developed to design high-performance fast-charging strategies in the presence of temperature and voltage constraints[11].

However, model-based approaches have been shown to be inefficient since they require extensive expertise knowledge and their performance is largely limited by the model complexity. Given the overwhelming complexity in modeling battery's physical and chemical properties with electrochemistry[12], data-driven approaches arise by treating the design of charging strategies as a black-box problem and using experimental data to estimate the optimal solution. Data-driven approaches are advantageous in that they do not require domain knowledge, instead, they only need experimental data to extract valuable insights and establish statistical machine learning models. For instance, Severson et al.[13] utilized the elastic net regression to predict battery cycle life based on the proactive charging and discharging features during early cycles. In addition, Hong et al.[14] proposed an approach to predict the remaining useful life (RUL) of batteries based on the voltage curve information from only a few early cycles, where dilated CNN is used to deal with the time-sequential voltage information and implicit sampling is adopted to measure the output uncertainty of the neural network.

A particular category of data-driven approaches directly utilizes experimental data for discovering the optimal charging strategies without explicitly building the data-driven battery models. For such black-box optimization methods, the charging phases are often split into divisions and the optimal charging current is determined for each division. Specifically, each charging strategy, consisting of a vector of candidate charging currents across these divisions, is evaluated by performing experiments and then the optimal strategy is obtained by comparing different charging strategies. A well-known

This work was supported by the National Natural Science Foundation of China (B. Jiang, Y. Wang and Z. Ma) under Grant 62273197, and the startup grant (Q. Lu) from the Texas Tech University. (*Corresponding author: Q. Lu*)

B. Jiang, Y. Wang, and Z. Ma are with the Department of Automation, Tsinghua University, Beijing, 100084, China (e-mail: bbjiang@tsinghua.edu.cn, yx-wang21@mails.tsinghua.edu.cn, mazhenghua94 @126.com).

Q. Lu is with the Department of Chemical Engineering, Texas Tech University, Lubbock, TX 79409, USA (e-mail: jay.lu@ttu.edu).

example is the grid search approach that performs a large number of repetitive experiments across different grid points (charging strategies) to yield an approximated optimum. However, this method becomes computationally and experimentally intractable if the number of decision variables is large and the experiments are expensive to perform[15]. Thus, it is desirable to acquire the solution for such black-box optimization with as few experiments as possible. To this end, Bayesian optimization (BO), a sequential decision-making method, has shown appealing performance for efficiently solving black-box optimization with much fewer experiments than grid search[16]. Research has been reported on using BO to tackle the design of charging strategies for batteries. For instance, Attia et al.[17] proposed a method to optimize fast-charging protocols, where a standard BO is utilized to maximize the battery cycle life under a fixed charging time. Jiang et al.[18] adopted BO to minimize the charging time where the convergence rates of several acquisition functions were compared. Jiang et al.[19] also studied the minimization of charging time, but with a focus on developing constrained acquisition function to ensure that the charging protocols respect the safety constraints of temperature and voltage.

Despite these impressive advancements, traditional BO approaches still suffer from limitations in introducing auxiliary information and characterizing uncertainty, particularly for designing charging strategies to optimize batteries' lifetime. In this work, we propose an improved BO method with the Bayesian recurrent neural network (BRNN) as the surrogate model. In this approach, a novel acquisition function is proposed that combines traditional expected improvement (EI) and upper confidence bound (UCB) criteria. Meanwhile, the auxiliary information, i.e., the charging voltages, is also included into the surrogate model for improving the prediction accuracy. We further compare our method with: (1) traditional BO method; and (2) BO with BRNN as the surrogate model but without auxiliary information. The advantages of the proposed BRNN-based BO charging approach is validated on the PETLION[20], a porous electrode theory-based battery simulator for examining the performance of fast charging protocols.

The rest of this article is as follows. The proposed BO method with BRNN as surrogate models is stated in Section II. The proposed RNN-based BO approach for optimizing battery charging strategies is elaborated in Section III. Simulation case studies are provided in Section IV for validating the developed methods, followed by conclusions in Section V.

## II. SURROGATE MODEL DESIGN FOR DEEP BO

A typical BO can be formulated as
$$\max_{x \in A} f(x) \quad (1)$$
where the feasible set $A$ is relatively large and the objective function $f(x)$ is expensive to evaluate. Generally, BO adopts a surrogate model (e.g., Gaussian process) to approximate the mapping from variable $x$ to the objective function $f(x)$, and constructs an appropriate acquisition function (AF) to decide the next sampling point during the iteration process.

### A. Recurrent Neural Network

In this research, the objective function of charging strategy optimization is to maximize the battery lifetime by finding the best multi-dimensional decision variables that consist of charging currents across different charging divisions. The most widely used surrogate model, Gaussian process (GP), can be used to predict battery lifetime based on charging currents. However, GP models only rely on predefined kernel functions to describe the relation between different charging currents with the distance information. Such models are too simple to reflect the complex electro-chemical correlations between charging currents and battery lifetime. Moreover, they cannot fully use a wide range of valuable auxiliary information generated from experiments such as charging voltages, temperature, and internal resistance. More importantly, GP may not capture the dynamic information from the time-series battery charging/discharging data. This motivates us to propose improved methods, particularly based on the recurrent neural network (RNN)[21], to enhance the prediction of battery lifetime. A unique feature of RNN is that the current output not only depends on the current input, but also on the previous inputs, which are transmitted in the form of hidden states. Since RNN can capture the temporal correlation in data series, it is widely used in sequential data processing scenarios such as natural language processing[22].

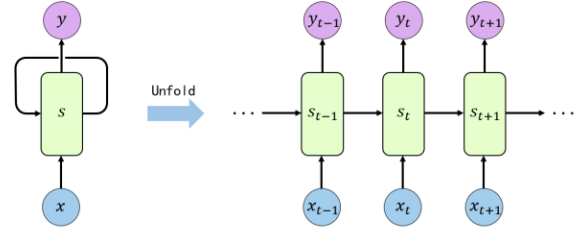

**Fig. 1.** The typical structure of a RNN[21].

As in Fig. 1, a typical RNN architecture consists of an input layer, one or more hidden layers, and an output layer[23]. Denote the input at time $t$ as $x_t$, the state as $s_t$, and the predicted output from RNN as $y_t$. The input layer maps the input $x_t$ to be combined with the current state $s_t$, which is then transitioned by the hidden layer to obtain the next state $s_{t+1}$. The predicted output $y_t$ is obtained by mapping the current state value $s_t$ via the output layer. The generic form for state transition and output prediction from RNN is
$$s_t = f_s(x_t, s_{t-1}|\theta_s), \quad y_t = W_y s_t + b_y \quad (2)$$
where $W_y$ and $b_y$ are the output layer parameters, and $\theta_s$ are the parameters in the input and hidden layers. A more explicit formulation of the state transition for RNN is shown as
$$s_t = \phi(W x_t + U s_{t-1} + b) \quad (3)$$
where $W$ is the input layer parameter, and $U$ and $b$ represent the hidden layer parameters. The function $\phi(\cdot)$ denotes the nonlinear activation function used by the network. In contrast, the output of a non-recurrent neural network only depends on the current input, without taking previous inputs into account. Specifically, the relation between input $x_t$, state $s_t$, and output $y_t$, for a non-recurrent network, is shown to be:
$$s_t = f_s(x_t|\theta_s), \quad y_t = W_y s_t + b_y \quad (4)$$
where the parameters are defined similarly to (2). Similar to the state-space model for dynamical systems, the state of RNN can capture the dynamical correlations and utilize it for predicting the output. In other words, the latent state serves as a bridge to 3connect past states/inputs with the prediction of outputs. However, traditional RNN as in Fig. 1 often cannot well capture long-term information[24]. To this end, the long short-term memory (LSTM) network, a variant of RNN, will be used in this work to alleviate this problem[25].

*B. Bayesian Recurrent Neural Network*

Although the RNN and LSTM networks are advantageous in capturing dynamical dependencies, they are unable to estimate prediction uncertainty due to their deterministic nature. Such statistical quantifications are critical for assessing the confidence level associated with the subsequential decision-making, particularly Bayesian optimization and inference. Therefore, Bayesian RNN has been developed to provide uncertainty information for the output prediction[26]. The work on Bayesian RNN was pioneered by Gal et al.[27] where variational dropout approach was adopted to obtain the Bayesian approximation of output prediction. Such techniques are further adopted and extended by Sun and Braatz[23] for fault detection and identification. For these methods, the full-scale Bayesian framework is utilized by treating all network parameters as random variables. The overall predicted output of the network is obtained by taking the expectation of individual predicted output under given parameters with respect to the posterior of these parameters. However, such methods involve an overly large size of parameters, posing significant challenge for computation. In this work, we adopt the strategy in Lázaro-Gredilla et al.[28] by only treating the parameters of the last layer of the RNN as random variables. This approach can not only avoid the computation issue associated with full-scale Bayesian setup, but also generate the distribution of the predicted output that is critical for the subsequent decision-making within the Bayesian optimization framework. The output of the RNN, under given network parameters, is assumed to be Gaussian distributed in this work.

We consider the typical BO setup in (1) with BRNN as a surrogate model to predict output $y_t$ (i.e., battery cycle life) under any given input $x_t$ (i.e., charging strategies). Denote $D$ as the set of data samples, and $\theta_y$ as the parameters of the last layer of the BRNN, which are assumed to be random. Then the posterior distribution of $y_t$, given the sample dataset $D$, is

$$p(y_t|D) = \int p(y_t|\theta_y) p(\theta_y|D) \, d\theta_y \tag{5}$$

where $\theta_y = [W_y \; b_y]$ and $p(\theta_y|D)$ is the posterior distribution of the random parameters in the last layer of BRNN under given dataset $D$, and $p(y_t|\theta_y)$ is the presumed distribution of the output under a fixed $\theta_y$. With such a Bayesian framework, the posterior distribution of output $y_t$ is essentially a weighted average of that under all possible parameter values where the weight is the posterior of parameter $\theta_y$ under dataset $D$. Given the posterior distribution $p(y_t|D)$, we denote $\mu_{y_t}(x_t)$ and $\sigma_{y_t}(x_t)$ as the mean and standard deviation of $y_t$ for any given input $x_t$. Then the AF can be constructed from these posterior statistics to determine the subsequential optimal input for attempt.

III. DEEP BO WITH BRNN FOR BATTERY FAST CHARGING

*A. Battery Fast Charging Optimization Problem*

The objective of this work is to discover the optimal fast charging strategy for lithium-ion batteries, with as few experiments as possible, that can slow down the battery degradation and respect safety constraints. The inherent conflict between battery aging and fast charging makes the problem of seeking the optimal charging strategy to be nontrivial. To simplify this problem, we fix the charging time to be a constant and design the charging strategy under this constant charging time, with the aim of reducing the battery aging and satisfying the safety constraints.

First, the battery cycle life is defined as the number of cycles when the battery capacity degrades to 80% of the nominal capacity. In general, longer cycle life indicates slower battery degradation. The charging protocols will follow the multi-constant-current-step charging profile[18, 29]. Specifically, we divide the fixed charging time to $n$ steps, with the charging current in the $i$th step defined as $I_{(i)}$. Then the aforementioned problem can be modeled as a combinational optimization:

$$\arg \max_{I_{(1)}, I_{(2)}, \cdots, I_{(n)}} y = f(I_{(1)}, I_{(2)}, \cdots, I_{(n)}) \tag{6}$$

where $y$ indicates the battery cycle life and $f(\cdot)$ represents the mapping from charging strategy to battery cycle life.

We specify the function $f(\cdot)$ in (6) as the objective function of BO. Since there is no analytical expression for $f(\cdot)$, we can only establish an approximate model $\hat{f}(\cdot)$ (i.e., the surrogate model) via experimental observation data. In addition, the design of acquisition function is also important for finding satisfactory charging strategies in the parameter space. This work mainly focuses on developing novel surrogate models utilizing additional battery experiment data to improve the battery lifetime prediction, as well as a novel acquisition function to efficiently find the optimal charging strategy. The proposed novel surrogate model and acquisition function will be discussed in the two subsections below.

*B. Battery Lifetime Prediction with LSTM Networks*

To design strategies for battery charging, the ultimate objective is to discover the optimal charging strategies that maximize the battery cycle life. To this end, we choose the battery cycle life as the output of the LSTM network and the charging currents across different charging steps as the input. However, directly predicting battery lifetime only based on charging currents is always difficult[13]. On the other hand, the voltage during the charging process is more indicative because it can better reflect the health status of a battery and thus can be used to speculate the battery lifetime. In addition, the voltage also has a strong relationship with the current since it is determined by the current during the charging process. Therefore, as an intermediate variable, voltage can serve as a bridge connecting the current and the battery lifetime.

However, different from charging currents, the voltage continuously changes over time and charging steps, and also varies slightly across different charging-discharging cycles. In addition, after completing all charging-discharging cycles, one can gather a large number of voltage measurements, and thus the voltage data throughout all cycles forms a high-dimensional vector. Such high-dimensional voltage information is hard to be incorporated into the LSTM network. Therefore, it is necessary to reduce its dimension, meaning to find a low-dimensional representation of the voltage data obtained from all experimental cycles. Given the observation that voltage curves from the first few cycles are similar, we choose the first cycle as the representative (see Fig. 2(a) and Fig. 2(c)). Next, we discard the irrelevant discharging and constant-voltage steps from the voltage curve because they are not affected by charging currents, as in the shaded area in Fig. 2(b) and Fig. 2(d). Then we divide the truncated voltage curve into several parts based on the state of charge (SOC), corresponding to different charging steps. For each part, we construct two or three statistical features to represent the original voltage curve to reduce the dimension. Here four schemes are used to produce statistical features:

**Scheme 1**: use the voltage mean and variance in each part.

**Scheme 2**: use the 20%, 50% and 80% quantile of the voltage in each part.

**Scheme 3**: fit the voltage curve by linear regression, then use the slope and intercept of the linear function in each part.

**Scheme 4**: fit the voltage curve by quadratic regression ($y = ax^2 + bx + c$), then use the parameters ($a, b, c$) in each part.

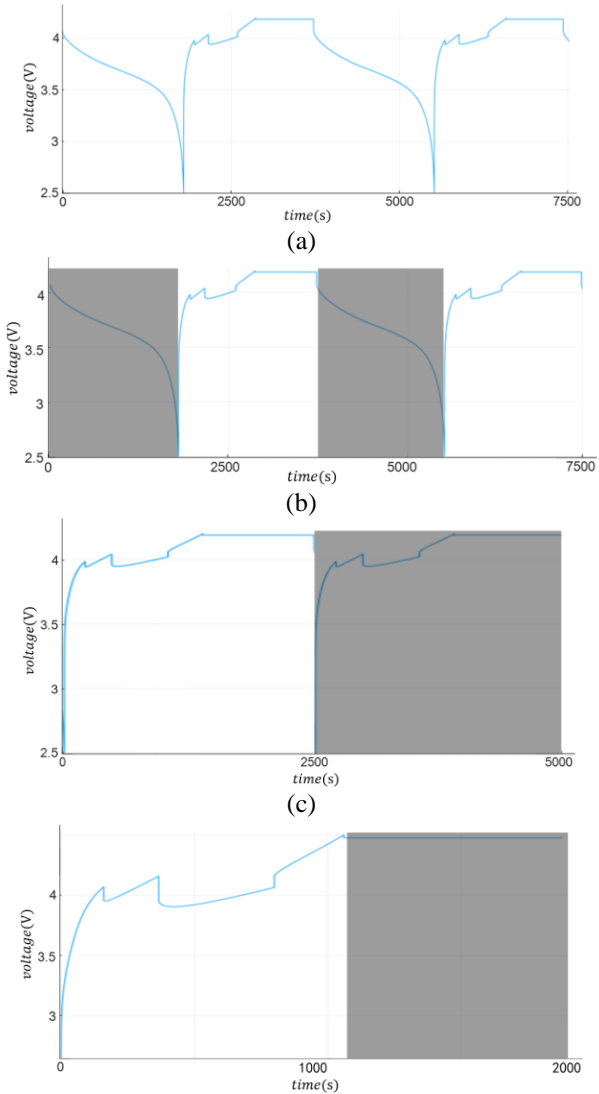

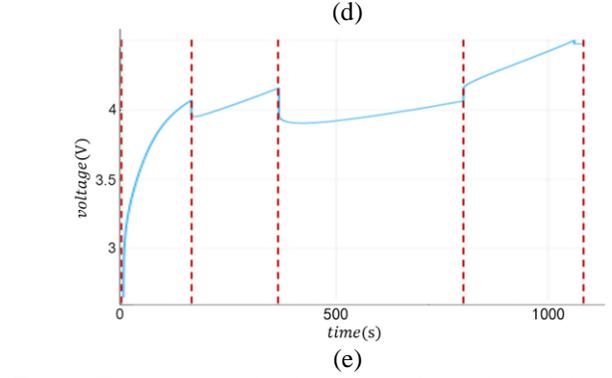

**Fig. 2.** The process of reducing the dimension of the voltage data. (a) The original voltage curve; (b) Omit discharging part; (c) Retain only one cycle; (d) Omit the CV step; (e) Separate the charging process based on the SOC.

The entire process to reduce the dimension of voltage data is illustrated in Fig. 2. Let us denote the charging currents as $I$, and the voltage features as $U$. In our approach, we divide the constant-current (CC) charging process into several steps, and the charging current in each step is defined as $I_1, I_2, \cdots, I_n$, where $n$ is the number of steps. The set $\{I_1, I_2, \cdots, I_n\}$ of currents forms the decision variables of the BO. Since the feature vector $U$ is obtained after the above dimension reduction of the voltage data, it is already a low-dimensional array containing the sequence information. Inspired by the role of hidden states in the RNN, in this work, we propose to use the low-order voltage features $U$ as intermediate states for enhancing the battery lifetime prediction with current input $I$, as shown in Fig. 3. Specifically, the relation between charging strategies $I$ (input), voltage features $U$ (state), and predicted battery cycle life $y$ (output) is depicted below:

$$\hat{U} = \hat{g}(I), \quad \hat{y} = \hat{h}(I, \hat{U}) \quad (7)$$

With this novel architecture, the prediction of battery lifetime becomes a two-step process. First, we use the charging strategy $I$ to predict the state $U$, followed by the second step of using the combined $U$ and $I$ to predict battery lifetime $y$. Note that combining $U$ and $I$ for predicting the battery lifetime is equivalent to adding an extra feature $U$ to supervise and improve the prediction of battery lifetime $y$ directly from $I$.

Another important fact is that strong temporal correlation exists within the charging current sequence $\{I_{(1)}, I_{(2)}, \cdots, I_{(n)}\}$, i.e., the present charging current can impact the final cycle life along with the subsequent charging currents throughout the entire cycle. Therefore, to capture such dynamical relation, we utilize an LSTM network to model the connections between charging currents across all steps. In other words, the $\hat{g}(\cdot)$ function above will be described by an LSTM network. For the lifetime prediction function $\hat{h}(\cdot)$, we will represent it by a traditional multi-layer perceptron (MLP) network that consumes the combined $U$ and $I$ as the inputs.

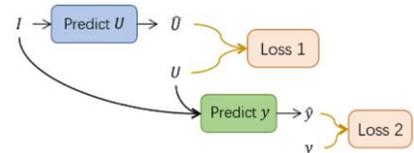

**Fig. 3.** The proposed framework to predict battery lifetime.

## C. Acquisition Function Design

After the previously established surrogate model, the next task for BO is to decide which point to sample in the subsequent iteration. Such decision making is often achieved by optimizing a constructed AF. Common strategies to establish AFs include expected improvement (EI), upper confidence bound (UCB), probability of improvement (PI), and Thompson Sampling (TS)[30]. The objective of the EI-based AF is to maximize the expected improvement

$$\alpha_{EI}(x) = \sigma_f(x)((Z_f(x)\Phi(Z_f(x))) + \phi(Z_f(x))) \quad (8)$$

with

$$Z_f(x) = \frac{f(x^+) - \mu_f(x)}{\sigma_f(x)} \quad (9)$$

where $\Phi(\cdot)$ and $\phi(\cdot)$ are the cumulative distribution function and the probability distribution function for normal distribution, respectively; $f(x^+)$ stands for the best observation value so far; $\mu_f(x)$ and $\sigma_f(x)$ indicate the posterior predictive mean and standard deviation for the objective function, respectively. The UCB-based AF has a simpler expression that is given by

$$\alpha_{UCB}(x) = \mu_t(x) + \kappa \sigma_t(x) \quad (10)$$

where parameter $\kappa$ balances posterior mean $\mu_t(x)$ and variance $\sigma_t(x)$.

In this work, we propose to use a hybrid AF for facilitating the BO. Initially, we use the EI-based AF to guide the search process. However, this method often becomes inefficient if the expected improvement of the entire parameter space is low, which means that the estimation of the objective function is inaccurate, causing the algorithm to fall into a local optimum. In this case, we can resolve this issue by further switching to the UCB-based AF. In practical applications, it often occurs that the maximum of the EI and the UCB falls on the same sample point[31]. A reasonable explanation is that both EI and the UCB criteria consider the uncertainty of the surrogate model when searching the subsequent sample point. In particular, the EI criterion calculates the possible improvement based on a normal probability distribution and the UCB criterion calculates the confidence bound based on the posterior variance. This often leads to the coinciding maximum for the EI- and the UCB-based AFs when the uncertainty exists[32]. Under such situations, the charging strategy corresponding to the UCB in a certain quantile (often between 75% to 90%) can be selected as the candidate strategy. If the selected charging strategy still repeats, then the final strategy will be randomly selected. When the maximum point of the EI has already been explored, a charging strategy corresponding to UCB in a certain quantile can make the algorithm jump out of the local optimum to enhance the exploration, while still taking the advantage of the existing prediction. Therefore, such a scheme allows us to make better usage of the existing learning results of the lifetime prediction network, to fully exploit areas with long lifetime expectation in the input space. This algorithm that combines EI and UCB can be described in Algorithm 1.

**Algorithm 1.** Deep Bayesian optimization with combined acquisition function.

1: Decide exploration leaning or exploitation leaning based on $D_y$ and rounds
2: **for** $I_i \in I$ **do**
3:   **if** exploitation leaning **then**
4:     $\sigma(\hat{y}_i) \leftarrow 0.5 \cdot \sigma(\hat{y}_i)$
5:   **end if**
6:   Compute $EI(\hat{y}_i)$
7: **end for**
8: **if** $\max_{I_i \in I} EI(\hat{y}_i) > 0.001$ **then**
9:   **return** $\arg\max_{I_i \in I} EI(\hat{y}_i)$
10: **end if**
11: **for** $I_i \in I$ **do**
12:   **if** exploitation leaning **then**
13:     $\sigma(\hat{y}_i) \leftarrow 0.5 \cdot \sigma(\hat{y}_i)$
14:   **end if**
15:   Compute $UCB(\hat{y}_i)$
16: **end for**
17: $I_c = \arg\max_{I_i \in I} UCB(\hat{y}_i)$
18: **if** $I_c \notin D_I$ **then**
19:   **return** $I_c$
20: **end if**
21: **if** exploitation leaning **then**
22:   Randomly choose $I_c$ s.t. $UCB(\hat{y}_c)$ is between 75% and 90% percentile of $UCB(\hat{y}_i)$, $I_i \in I$
23: **else**
24:   Randomly choose $I_c$ s.t. $UCB(\hat{y}_c)$ is between 85% and 100% percentile of $UCB(\hat{y}_i)$, $I_i \in I$
25: **end if**
26: **if** $I_c \notin D_I$ **then**
27:   **return** $I_c$
28: **end if**
29: **while** $I_c \in D_I$ **do**
30:   Randomly choose $I_c$ from $I$
31: **end while**
32: **return** $I_c$

## IV. RESULTS

In this section, we first introduce the used battery simulator for validating our methods. Then, we apply our proposed BRNN-based lifetime prediction method and compare it to the widely used GP-based method. Finally, we test the performance of the proposed BO with mixed EI-UCB-based acquisition function for fast charging protocol optimization to demonstrate the efficacy of using the hybrid AF.

### A. Simulation

The simulation is implemented on the PETLION[20], a porous electrode theory-based battery simulator. We split the charging current into three steps: $I_{(1)}$, $I_{(2)}$, and $I_{(3)}$, where $I_{(1)}$ and $I_{(2)}$ are free variables and $I_{(3)}$ is determined by the former currents. Define $\Delta Q_{(i)}, i = 1,2,3$, as the change of SOC before and after the $i$-th charging step, and denote $\Delta t_{(i)}, i = 1,2,3$, as the duration of the three steps. Then it can be obtained as

$$t_{(i)} = \frac{\Delta Q_{(i)}}{I_{(i)}}, i = 1,2, \quad t_{(3)} = t_f - t_{(1)} - t_{(2)}, \quad I_{(3)} = \frac{\Delta Q_{(3)}}{t_{(3)}} \quad (11)$$

where, $t_f$ is the total charging time, and $t_f = 800$ s is used in this work. $I_{(1)}$, $I_{(2)}$, and $I_{(3)}$ are the charging currents corresponding to charging battery from the SOC of 0% to 20%, from 20% to 40%, and from 40% to 80%, respectively. Both $I_{(1)}$ and $I_{(2)}$ range from 2.2 C to 6.0 C in this work, with step length of 0.1 C. We split the ranges of $I_{(1)}$ and $I_{(2)}$ into fine grids and simulate the battery lifetime at each grid. The battery lifetime over the entire 2D space across different current values is shown in Fig. 4.

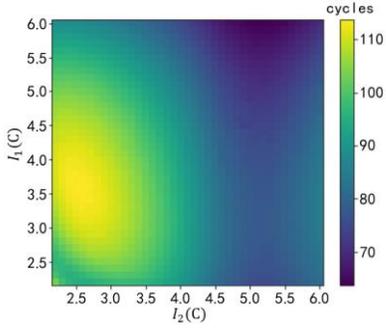

**Fig. 4.** Illustration of the simulated battery lifetime across the 2D parameter space.

*B. Surrogate Models for Lifetime Prediction*

In this section, we are to validate the predictive ability of the BRNN-based surrogate model. For comparison, we use both non-recurrent network and BRNN to predict the battery lifetime, where the inputs to both networks are identical, including charging currents and charging voltages. Fig. 5 shows the prediction result of the non-recurrent network. Comparing with Fig. 4, all plots in Fig. 5 are quite different from the true plot, regardless of the scheme that is employed. This indicates the incapability of using non-recurrent network as surrogate models for predicting battery lifetime.

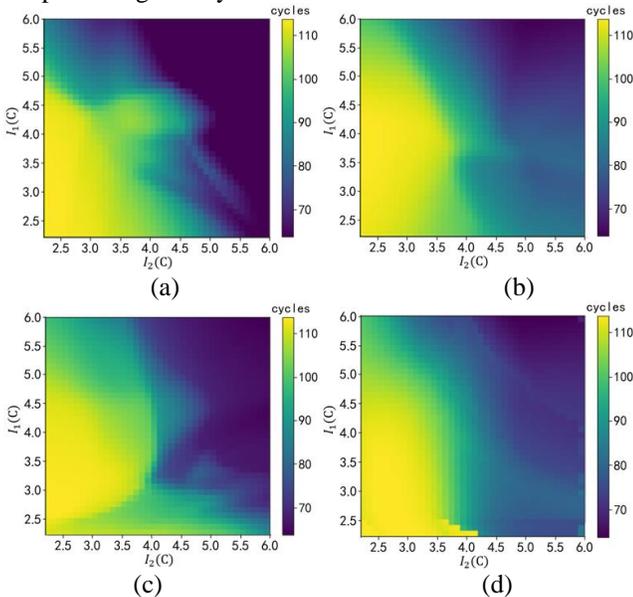

**Fig. 5.** Battery lifetime prediction by the non-recurrent network with both charging currents and voltages as inputs. Each plot (a), (b), (c) and (d) indicates Scheme 1 to 4.

In contrast, the prediction results of the BRNN-based surrogate model show much improved accuracy as in Fig. 6. With this surrogate model, all of these four schemes can yield precise predictions of battery lifetime and in particular, Scheme 3 and Scheme 4 present better performance than Scheme 1 and 2. Therefore, the following section will only employ Scheme 3 for demonstrating the proposed method and its comparison with other methods.

*C. Bayesian Optimization of Fast Charging Protocols*

We choose four random charging strategies to initialize the BO process. With these starting points, BO is then conducted to sequentially search the optimal strategy, until reaching the given maximum number of iterations, i.e., 20 rounds. Since the optimization result is affected by the initializing points, we repeatedly run the entire optimization routine by 20 times, and then calculate the mean result as the final result.

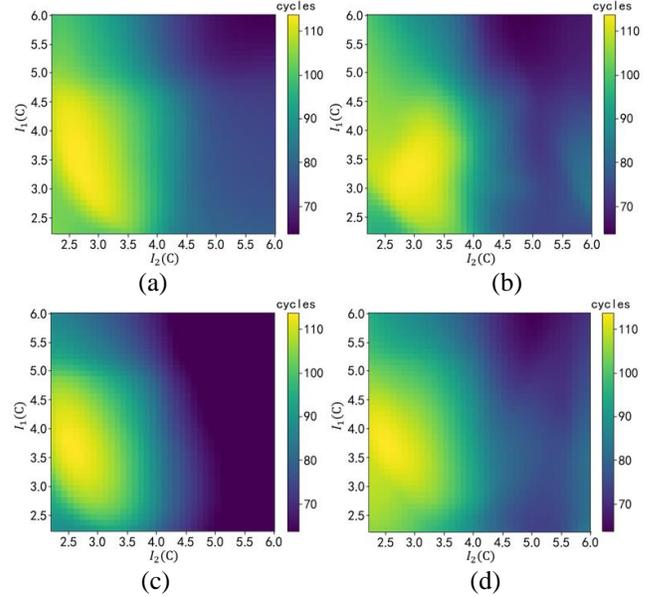

**Fig. 6.** Battery lifetime prediction by BRNN network with both charging currents and voltages as inputs. Each plot (a), (b), (c) and (d) indicates Scheme 1 to 4, respectively.

For comparison, we employ three variants of BO: defined respectively as the baseline, control group, and the proposed method. First, the baseline method adopts the traditional BO with GP as the surrogate model and the EI criterion as the AF. Second, the control group method uses BRNN as the surrogate model and the EI criterion as the AF. However, the BRNN here only uses the charging currents as the input without using the charging voltages. Finally, the proposed method employs BRNN as the surrogate model and mixed EI and UCB criteria as the AF, with both charging currents and charging voltages utilized as the inputs to the BRNN.

*1) The baseline method*

For this method, the prediction result of the surrogate model after 20 iterations based on GP and the trajectory of the sample points over the BO iterations are shown in Fig. 7. We can see that the surrogate model cannot provide an accurate prediction for the battery lifetime even after 20 rounds of iterations. Meanwhile, the selected samples points by BO are largely dispersed in parameter space without converging to the optimum. This phenomenon indicates that the traditional BO, based purely on the GP as the surrogate, cannot fully learn the parameter space, and thus is not able to discover the optimum rapidly. Such an observation, on the other hand, underpins the importance of developing advanced surrogate models with more complex machine learning models and additional battery features for enhancing the battery lifetime prediction.

*2) The control group*

The control group adopts the BRNN as the surrogate model, but the charging voltages are not considered as the input to the network. The left plot of Fig. 8 demonstrates the prediction result of the surrogate model after 20 BO iterations. It is observed that after these iterations, the surrogate model can discover the region where the global optimum resides (see the yellow region), despite that the predicted performance over the entire region still has a large discrepancy with respect to the true values as in Fig. 4. The trajectory of sample points from the control group is shown in the right plot of Fig. 8. It reveals that the optimization algorithm often visits the edge of the parameter space, where the battery lifetime is far less than the optimal value. Although the sample points finally converge into the region of the optimum, the control group still suffers from low efficiency in convergence.

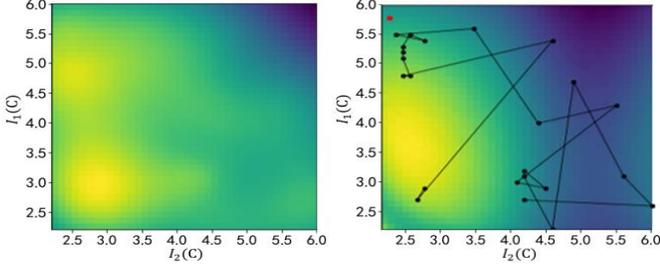

**Fig. 7.** Left: The prediction result of the surrogate model after 20 iterations based on the baseline method; Right: The trajectory of the sample points over BO iterations, where the background is the simulation result from Fig. 4, and the red dot stands for the determined next sample points.

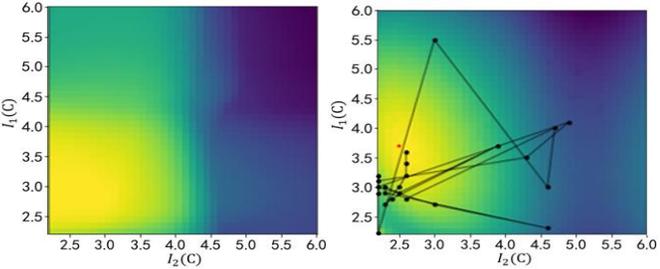

**Fig. 8.** Left: The prediction result of the surrogate model of the control group after 20 BO iterations; Right: The trajectory of the sample points over BO iterations based on the control group. The red dot is the determined next sample point.

*3) The proposed method*

The proposed method uses the BRNN as the surrogate model, and both charging currents and voltages are employed as the input to the network. The prediction result of the surrogate model and the trajectory of the sample points during BO iterations are shown in Fig. 9, respectively. The left plot of Fig. 9 shows that after 20 BO iterations, the proposed method can predict the battery lifetime with a high accuracy, as indicated by the high similarity between the color maps in Fig. 9 and Fig. 4. In fact, for the proposed method, the surrogate model can capture the true model rapidly. Fig. 10 shows the estimated posterior distribution of the battery lifetime based on our proposed method at iterations 1, 5, 10, and 15. One can see despite the large difference from the true distribution at the beginning, the proposed surrogate model can well capture to the true model only after 10 iterations. This verifies the effectiveness of our BRNN-based surrogate model with both currents and voltages as inputs. Fig. 9 (right) shows the trajectory of the sample points over BO iterations based on the proposed method. One can see that the sample points can converge to the neighborhood of the optimum after a few iterations although the starting points are far from the optimum. Thus, our method exhibits superior performance in both optimization result and computation efficiency.

*4) Comparison of Loss Function*

To quantitatively compare the prediction performance of the surrogate models, we define the following loss function:

$$L = y^* - \max(D_y) \quad (12)$$

where $y^*$ denotes the global optimal battery lifetime in the simulation dataset, and $D_y$ denotes the set of all the estimated data points. We run the optimization 20 times with the three candidate methods, and then calculate the mean and variance of the loss functions (12).

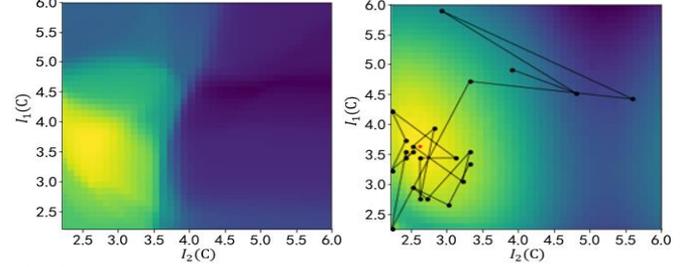

**Fig. 9.** Left: The prediction result of the BRNN-based surrogate model after 20 iterations; Right: The trajectory of the sample points over BO iterations based on the proposed method. The red dot is the determined next sample point.

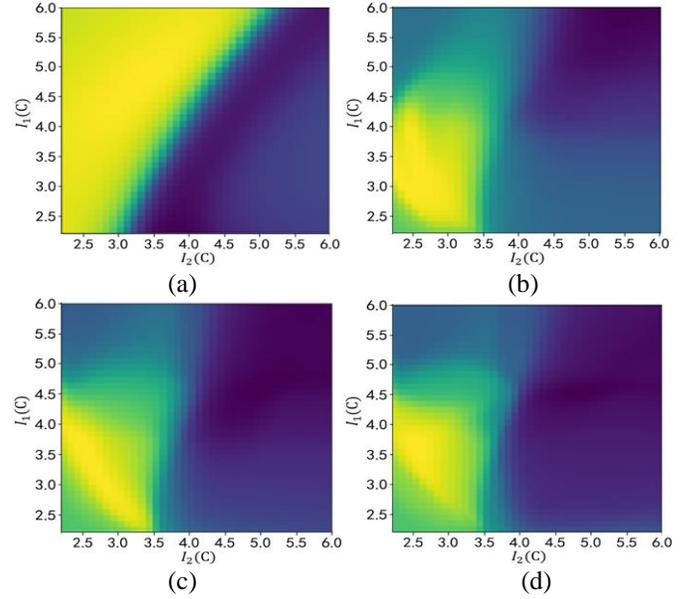

**Fig. 10.** The predicted battery lifetime based on the proposed BRNN surrogate model over BO iterations. Each plot (a), (b), (c) and (d) indicates iterations 1, 5, 10, and 15, respectively.

As in Fig. 11, the proposed method has the fastest downtrend and the lowest loss value after convergence. For the robustness of these algorithms, i.e., considering the value of the mean plus the variance of the loss, we can see that all algorithms have a high loss value at the beginning of the optimization. However, the loss value of the baseline method remains high whereas that of the proposed method decreases rapidly. In addition, the loss value of our method also clearly outperforms that of the control group, as indicated by the large difference between the blue and green curves over the last few iterations. In summary, the discovered solution and the optimization efficiency of the proposed method surpass those of the control group and the baseline method. This observation indicates that adopting BRNN as the surrogate model and usings charging voltages as extra information can significantly improve the BO for designing optimal charging strategies.

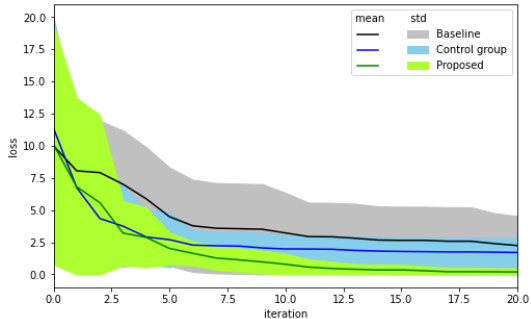

**Fig. 11.** The loss function curves of the baseline (grey), the control group (blue), and the proposed method (green).

## V. Conclusion

In this article, a novel Bayesian optimization algorithm is proposed with the BRNN as the surrogate model and the combination of EI and UCB criteria as the acquisition function for fast charging strategy optimization. The proposed optimization method was validated using the data generated by the PETLION simulator. The results show that the BRNN-based surrogate model can predict the battery lifetime with a higher accuracy than the widely used BO based on GP or without utilizing additional voltage information. Simulation results further show that the proposed method also has the fastest convergence speed for fast charging optimization. The proposed deep Bayesian optimization approach embodies the potential of using data-driven methods to handle complicated problems of battery parameter optimization, such as the design of next-generation electrode and electrolyte chemistries[33].